\documentclass{aa}
\usepackage{graphicx}
\begin{document}

   \title{Reconstruction of the microlensing light curves of the Einstein Cross, QSO2237+0305: possible evidence of an accretion disk with a central hole}

   \subtitle{Microlensing HAE analysis for the OGLE \& GLITP monitoring data}

   \author{Dong-Wook Lee\inst{1}, J. Surdej\inst{2}, O. Moreau\inst{2}, C. Libbrecht\inst{2} \and J.-F. Claeskens\inst{2}
          }

   \offprints{D.-W. Lee: macrolensing@yahoo.co.kr}

   \institute{Astrophysical Research Center for the Structure and Evolution of the Cosmos (ARCSEC), 
              Sejong university, Goonja-dong, Kwangjin-gu, Seoul, Republic of Korea \\
         \and
              Institute of Astrophysics and Geophysics, University of Li\`ege,
              All\'ee du 6 ao\^ut 17, B5c, B-4000 SartTilman (Li\`ege), Belgium \\
             }

   \date{Submitted to A\&A}

   \abstract{ Recent OGLE (Optical Gravitational Lensing Experiment) and GLITP (Gravitational Lens International Time Project) monitoring data for QSO 2237+0305 (Huchra et al. \cite{huchra}) have been analyzed 
              through a newly optimized N-body microlensing analysis method, \emph{the Local HAE Caustic Modeling (LOHCAM)}.
              This method simultaneously solves for the size of the source and N-body HAE (High Amplification Events) caustic shapes 
              in the source plane and determines those sizes only as a function of the projected transverse velocity of the source.
              By applying this method to the light curves of the A \& C lensed components in the Einstein Cross, these data are accurately reconstructed for the first time. 
              From these modeling studies, we report several interesting results: the minimum number of microlenses required for possible caustic models, the possible evidence of an accretion disk with a central hole at the heart of the quasar, the size of the UV-continuum source, the masses of the microlenses being directly responsible for the observed HAEs, the estimated mass range of a super massive black hole (SMBH) in QSO 2237+0305 and finally some clues for the direction of the source motion in the sky.

   \keywords{quasars: individual (2237+0305) --
               gravitational microlensing --
               accretion disk  
               }
   }
 
  \authorrunning{D.W.Lee et al}
  \titlerunning{Possible evidence of an accretion disk with a central hole in QSO 2237+0305}
  \maketitle
%

\section{Introduction}

Gravitational microlensing effects induced by compact objects in a galaxy can be used as a cosmic microscope to probe at micro-arcsec angular scales the structure of the lensed background quasar (Chang \& Refsdal \cite{chang}, Young \cite{young}, Chang \& Refsdal \cite{chang84}, Kayser, Refsdal \& Stabell \cite{kayser}). A few years ago, the OGLE team (Wozniak et al. \cite{wozniak}) has reported impressive microlensing light curves for the 4-macro images (hereafter denoted A, B, C \& D) of QSO 2237+0305 ($z\sim1.695$) lensed by a nearby spiral galaxy ($z\sim0.0394$). This system constitutes a very promising candidate for microlensing studies (Kayser, Refsdal \& Stabell \cite{kayser}, Kayser \& Refsdal \cite{kayser89}). Indeed, because the expected time delays are so short, about one day (Schneider et al. \cite{sch}), intrinsic variability should show up almost simultaneously in the four images so that any brightness variations affecting only some of the four single images are necessarily due to microlensing. Also, because of the large distance ratio between the QSO and the lens, microlensing should lead to more frequent and rapid high amplification events (HAEs) than for any other known gravitational lens system. 
So far, there have been several monitoring observations for this unique gravitational lens system, however, the OGLE data (Wozniak et al. \cite{wozniak}) are unique in that they cover prominent HAEs with unprecedented daily time resolution. Recent GLITP-ISIS data (Alcalde et al. \cite{alcalde}, Moreau et al. \cite{moreau}) provide some additional data around the microlensing peak of A and C.
 In section 2, we present data handling for the OGLE and GLITP data. In section 3, the LOHCAM technique is introduced. In section 4, we present the LOHCAM results and discussion.

   \begin{figure*}[ht]
   \centering
   \includegraphics{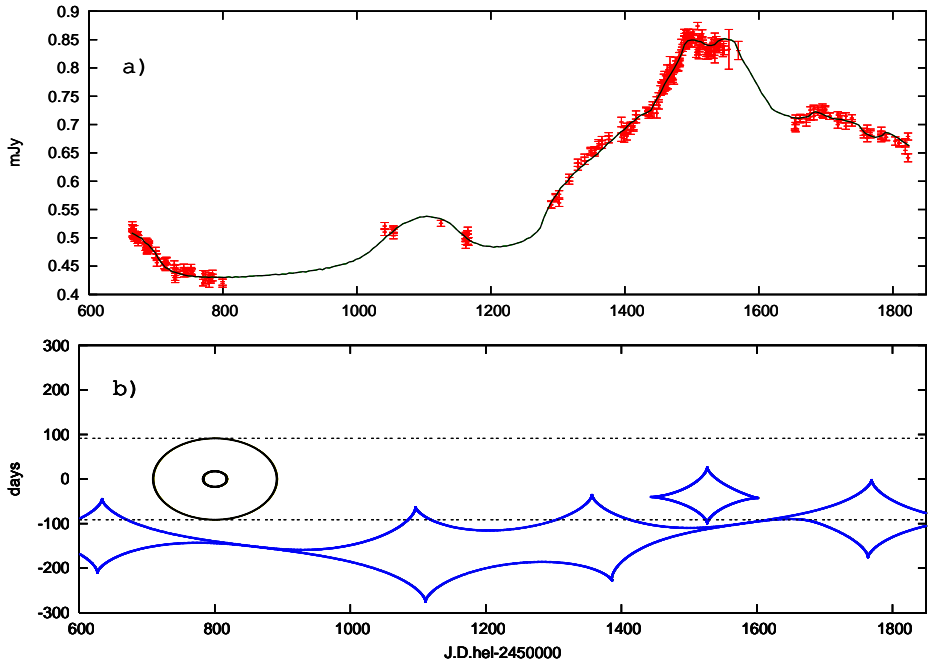}
   \caption{
    OGLE+GLITP data and model fitting results. Combined V-band light curves for image A and the possible caustic model. (a) The OGLE (231 points) \& GLITP-ISIS (53 points; around day 1500) data are shown altogether with the simulated light curve (solid line) using a disk with a central hole source model. Note the smooth and fine fitting over the entire data points, especially around the HAE peak (see more details in Fig. 3).  (b) The corresponding HAE caustic network (solid line) is shown with the straight track (dotted lines) for the moving source (Y=0). The concentric circles represent the sizes of the best-fitted inner and outer radii of the face on accretion disk. The sizes of the caustic network and of the source are represented in time units (days) along the X and Y axes.
    }
    \label{FigAAA}%
    \end{figure*}


\section{Data handling}
The OGLE data cover approximately 3 years with 231 data points, and the GLITP data cover 120 days with 53 data points. We note that two different methods have been applied to GLITP data analysis: GLITP-PSF fitting data \& GLITP-ISIS data (Alcalde et al. \cite{alcalde}, Moreau et al. \cite{moreau}). In this study, we use the OGLE data and GLITP-ISIS data which provided smaller error bars compared to GLITP-PSF fitting data. In order to combine these two data sets, first we corrected a zeropoint difference of 0.065 and 0.13 mag. in the V band between the OGLE and GLITP-ISIS data of A \& C, respectively, so as to match the slope of the OGLE data. Then, the original logarithmic magnitude scale is transformed into a linear flux unit (mJy) based on the same formulation given by the OGLE team. Figures 1a and 2a illustrate the resulting combined light curves in mJy unit that we have analyzed.

\section{Local HAE Caustic Model}

   \begin{figure*}[ht]
   \centering
   \includegraphics{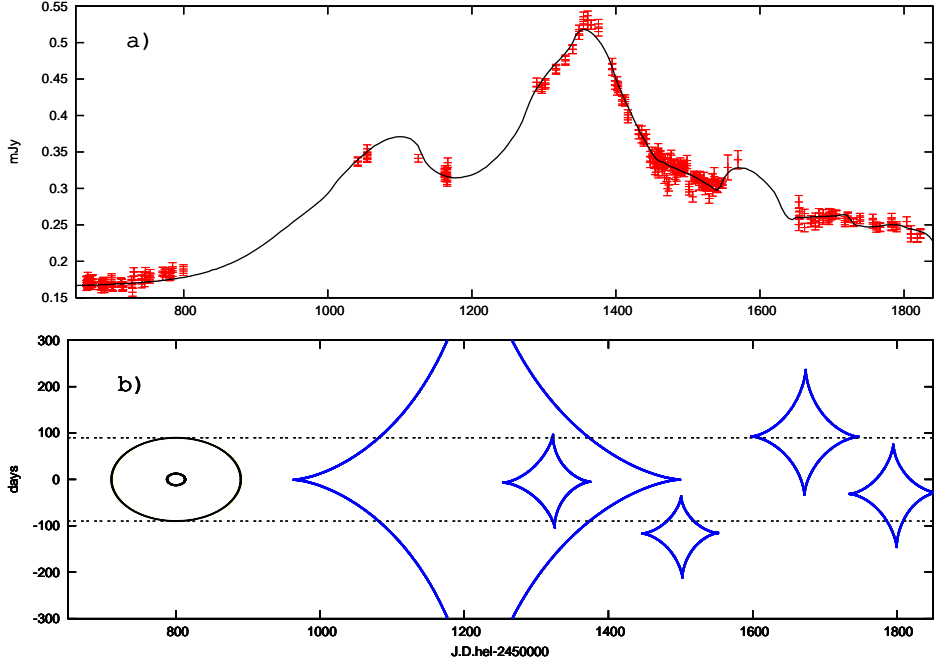}
   \caption{
    OGLE+GLITP data and model fitting results. Combined V-band light curves for image C and the possible caustic model. (a) The OGLE (231 points) \& GLITP-ISIS (53 points; around day 1500) data are shown altogether with the simulated light curve (solid line) using a disk with a central hole source model. (b) The corresponding HAE caustic network (solid line) is shown with the straight track (dotted lines) for the moving source (Y=0). The concentric circles represent the sizes of the best-fitted inner and outer radii of the face on accretion disk. The sizes of the caustic network and of the source are represented in time units (days) along the X and Y axes.
    }
    \label{FigCCC}%
    \end{figure*}


Usefulness of lightcurves of quasar microlensing is studied and discussed by many authors (Wambsganss Paczynski \& Schneider \cite{wamb}, Agol \& Krolik \cite{agol}\ and Mineshige \& Yonehara \cite{min}). They have investigated various methods to reconstruct spatial structure of quasar accretion disk. Recently, several authors (Yonehara \cite{yonehara}, Shalyapin \cite{shal}, Shalyapin et al. \cite{shal02}) have attempted to fit parts of the OGLE or GLITP data using HAE approximation formulae for a fold caustic and a cusp under the assumption of a single Chang-Refsdal microlens (e.g. Yonehara \cite{yonehara}) or a straight-line caustic (e.g. Shalyapin \cite{shal}, Shalyapin et al. \cite{shal02}). However, microlensing amplification maps (Wambsganss, Paczynski \& Schneider \cite{wamb}) even for component A, which is expected to have the smallest optical depth among the 4-images, suggest the possibility of a substantial amount of caustic clustering effects. Therefore, a single microlens or almost straight-line fold caustic model may not be adequate to interpret the complex HAEs observed for the Einstein Cross. We have developed a consistent Local HAE Caustic Model (LOHCAM; Lee \cite{lee3}) based upon N-body microlenses using a ray shooting method (Kayser, Refsdal \& Stabell \cite{kayser}) which simultaneously takes into account the extended nature of the background source and the shear effects due to the macro deflector (e.g. lensing galaxy). LOHCAM implicitly assumes that the HAEs are produced by those caustics located near the source and that the additional background amplification due to the non-local microlenses may be considered constant and negligible during a local HAE; it makes use of the fact that the microlensing amplification in the rayshooting method depends only on the size of the source and on the strength of caustics, regardless of the unlensed flux of the source.

For source models, we restrict ourselves to consider only two simple models: uniformly bright circular disk with a hole and a uniformly bright disk models assuming face on status. According to the traditional unified theory of AGN, we can naturally imagine the presence of a central hole in an accretion disk (i.e. an accretion disk with a central hole; see Fig. 2 Thorne \& Price \cite{thorne}) with a SMBH in the heart of a quasar. The free parameters in LOHCAM are the mass and coordinates of the local microlenses, the inner and outer radii of a circular source assumed to be a face-on accretion disk with a hole surrounding a central SMBH and the background flux level; i.e. 3N + 3 parameters. Additional pre-fixed parameters are the external shear, the shear direction angle defined with respect to the relative direction of the source motion and the number N of local microlenses.

The unknown minimum background amplification is assumed to correspond to the minimum flux levels present in each of the A (see Fig. 1a around day 800) and C (see Fig. 2a around day 700) light curves. These levels are also found to be representative of minima recorded in the compilation of past data (Wyithe, Webster \& Turner \cite{wyithe}).  We adopt representative shear values of 0.42 and 0.62 at A \& C from a general macro lens model for the case of the potential exponent ($\beta=1$) (Witt, Mao \& Schechter \cite{witt}), though the macrolensing parameters (e.g. $\sigma$, $\gamma$ ) are somewhat dependent on the adopted models (see also, Witt \& Mao \cite{witt94}, Wambsganss \& Paczynski \cite{wamb94}). We also note that although Schmidt, Webster \& Lewis (\cite{schmidt}) have reconstructed the macrolens model for the lensing galaxy of QSO2237 by using the observed light distribution (e.g. a bulge and bars), their model could not take properly into account the dark halo effect. Then, in order to take into account the dark halo of the lensing galaxy, we have chosen the macrolens model parameters from the most general model, the isothermal lens model ($\beta=1$). 

Based on the normalized lens equation (Kayser, Refsdal \& Stabell \cite{kayser}, Paczynski \cite{pacz}),

\begin{equation}
\vec{\eta} = \left( \begin{array}{cc}
               1+\gamma & 0   \\
               0   & 1-\gamma \\
              \end{array}\right)
\vec{\xi} - \sum m_{i}\frac{\vec{\xi}-\vec{\xi_i}}{|\vec{\xi}-\vec{\xi_i}|^2},
\end{equation}

where $\gamma$ is the external shear, $m_i$ represents the normalized mass of star i, $\vec{\eta}$, $\vec{\xi}$ and $\vec{\xi_i}$ the normalized co-ordinates in the source and the lens plane, respectively; the resulting microlensed lightcurves are then calculated. For simplicity, we ignore the totally unknown smooth dark matter term ($\sigma_c$); this could slightly affect some of the scaling relations (Kayser, Refsdal \& Stabell \cite{kayser}, Paczynski \cite{pacz}). Furthermore, due to the weak flux variation seen in the D light curve (Wozniak et al. \cite{wozniak}) of QSO 2237+0305, we assume that the intrinsic flux variations of the quasar are negligible and that all flux variations are due to microlensing effects during the monitoring campaigns. 

When fitting the light curves, we tried to find the minimum number N of local HAE microlenses in order to reproduce the whole light curves and to interpret the light curves as smooth as possible where data are not available. To convert the observed timescale into dimensionless normalized units for simulations, arbitrary time normalization is applied to the light curves. (e.g. 100 days). From the observed data, especially those of A, we notice that the average time scale between 2-consecutive turning points in the light curve is about 100 days. This fact may imply an important clue for the sizes of the source and caustics. However, the exact value for the time normalization is not important, because it produces the same results only if the initial conditions are same in absolute scale (see Eq. 2). By introducing a time normalization, the observed light curves transform into normalized (dimensionless) scale light curves. Then we can make use of the dimensionless normalized microlensing simulations compared with the normalized light curves without considering any cosmological effects. 

Of course, when we convert a normalized unit into a real physical unit, then we should take into account cosmological effects. A large number of preliminary simulations were then performed to identify plausible ranges for the initial parameters. Starting from the binary lens plus external shear cases (Lee \cite{lee}, Lee, Chang \& Kim \cite{lee2}), we have searched various caustic cuts to match the observed light curves. However, we could not find any satisfactory results from the binary lens model, so we increased the number of microlenses to find the possible minimum number caustic solutions compatible with the observational data. Finally, we found that the minimum number of microlenses needed to account for the observed HAEs is 5 for both images A \& C (Figs. 1b \& 2b).

Next, applying a simultaneous minimization method, the downhill simplex algorithm (Press et al. \cite{press}), the best chi-squared set of parameters characterizing the observational data was finally derived. It is important to note that knowledge of the apparent source velocity with respect to the caustics directly determines the absolute sizes of the source and caustics and, consequently, the mass of each microlens through the time normalization of LOHCAM. Through the LOHCAM analysis, we can directly derive the physical mass (M) of a microlens based on the normalized mass as follows;

\begin{equation}
  M=\frac{c^2}{4G}\frac{D_L}{D_S D_{LS}}\times(nm \times 100days \times V_T)^2
\end{equation}

where $c$ is the speed of light, $G$ is the gravitational constant, $D_L, D_{LS} \& D_S$ represent angular diameter distances, $nm$ represents the normalized mass, 100days represents the basic normalization unit in our simulations, $V_T$ is the projected transverse velocity of the source with respect to the caustics in the source plane measured by an observer.
%
%
   \begin{figure}[ht]
   \centering
   \includegraphics[width=8.5cm]{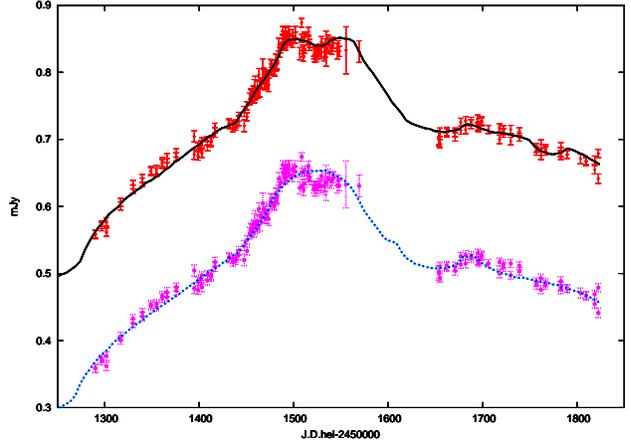}
      \caption{The difference between two source models for the A light curve.
               The upper graph shows the case optimized for a disk with a hole source model ($\chi^2=1.72$).
               The lower graph shows the case optimized for a disk source model  (slightly offset for clarity; $\chi^2=2.38$). 
               Note the much better matches for the case of the disk with a hole model for the double peak, turning points and a small dip.
               These two curves are the best-optimized cases by using a least number of microlenses (i.e. 5-microlenses), respectively.        
      }
         \label{FigAAD}
   \end{figure}
%


   \begin{table*}[htb]
    \begin{center}
      \caption[]{The best optimized model fitting results through LOHCAM.}
         \label{model}
    
          \begin{tabular}{cccccc}
           \hline
           \hline
                 & A & C \\
           \hline
            Inner source radius (($V_T/25,000km s^{-1})AU$) & 254 (+150/-119) & $\leq 349^{\mathrm{a}}$ \\
            Outer source radius (($V_T/25,000km s^{-1})AU$) & 1318 (+23/-73) & 1297 (+144/-77)  \\
            Masses ($10^{-2}\times(V_T/25,000km s^{-1})^{2} M_\odot$) & 1.07, 1.91, 1.19, 1.23, 1.29 & 102.02, 2.04, 1.20, 0.75, 0.34   \\
            Reduced $\chi^{2}$ & 1.72 & 1.86   \\
           \hline
          \end{tabular}
    
\begin{list}{}{}
\item[$^{\mathrm{a}}$] The inner source radius for the case of C is not well constrained. This may be due to the lack of data around the peak and the fact that the peak of HAE of C image is already affected by another large caustic (see Fig. 2). Quoted errors for the source sizes are $1 \sigma$. The errors were calculated from the 2-D (i.e. inner source radius versus outer source radius) $\Delta \chi^2$ contour map corresponding to $1 \sigma$ level, while fixing the other parameters at their best-fitted values. Please note that these values are functions of the transverse velocity and these values are derived assuming $V_T = 25,000 km s^{-1}$ in the source plane.
\end{list}
\end{center}
\end{table*}

\section{Results}

Although it is extremely hard to prove the uniqueness of our proposed caustic models as we never observe the caustics, we could derive some interesting and consistent results for the microlensing effects in the Einstein Cross. From the numerical simulations, we find that only a source model with a hole inside may easily account for the double peak observed in the light curve of component A (Figs. 1a \& b). Without such a hole, the simulated light curves fail to reproduce this double peak (Fig. 3). Even if we include additional microlenses (6 or 7 in total), we failed to find any good fitting results for the double peak with a disk source without a hole. We note that this double peak modeling for A is very similar to Grieger et al.(\cite {greiger})'s theoretical prediction for the case of an accretion disk with a central hole source. This result thus provides the first possible evidence of an accretion disk with a central hole for the far-UV continuum source of a distant quasar, namely QSO 2237+0305, observed at a redshift z=1.695 in the V band. In addition, our modeling produces a remarkably good fitting over the entire data points (Figs. 1a \& 2a), not just over selected parts of HAEs (Yonehara \cite{yonehara}, Shalyapin \cite{shal}, Shalyapin et al. \cite{shal02}). Our results are directly understood from the constant motion of the face on accretion disk source along a straight track across the caustic networks (Figs. 1 \& 2). The final $\chi^2$ fitting values are small, but not sufficiently small. This may be induced by our simple uniform brightness modeling for the surface brightness distribution of the source. As we can see from the fitting results, most of parts of the A data are well reproduced by the LOHCAM. Though the double peak modeling is evident, at the double peak region there is a slight misalignment between our modeling and the data; we need to make more sharp peaks. If some one can take into account more realistic surface brightness distribution of the source with our caustic and source model, the corresponding $\chi^2$ results should be near to unity because most of $\chi^2$ loss occurs at the peak region in our modeling. Shalyapin et al. (\cite{shal02}) made a HAE fitting for the light curve of A only based on GLITP-PSF data and derived a small $\chi^2$. However, we note that GLITP-PSF data has larger error bars compared to GLITP-ISIS data; so it is easy to have a smaller $\chi^2$ value. Moreover, unlike any other HAE studies made so far, our model fitting produces independent and similar solutions for the source size from both the A \& C light curves (Table 1). In our opinion, this testifies the coherence of the LOHCAM results, also suggesting that the contribution of the unknown smooth dark matter is not sufficient to produce over-focused cases in this lens system (Kayser, Refsdal \& Stabell \cite{kayser}).
 
From the best fitting results, we may now infer the physical sizes for the accretion disk and the central hole of the quasar and the masses of the microlenses. Please notice that all the results of LOHCAM only depend on the transverse velocity of the source with respect of caustics in the source plane measured by an observer; the size of source is linearly proportional to the transverse velocity and the masses of microlenses are proportional to the square of the transverse velocity. Although Kayser \& Refsdal (\cite {kayser89}) estimated $ 6,000 kms^{-1} $ for the projected transverse velocity of the source with respect to the caustics in the source plane, recently Kochanek (\cite{koch}) calculated a more realistic transverse velocity of QSO2237 in the source plane based on statistical analysis of the OGLE light curves. His estimation is roughly between $ 10,000 \sim 40,000 kms^{-1} $. Adopting a mean value ($ V_T=25,000kms^{-1} $) of his estimation, we estimate the size of source and the masses of microlenses (Table 1).

For the cosmological model, we assume $ H_o = 65 km s^{-1} Mpc^{-1} $ for the Hubble parameter and a flat Universe with a matter density $ \Omega_m = 0.3 $. The best-fitted outer and inner radii of the source for A are then found to be $ 1318 (+23/-73)\times(V_T / 25,000 km s^{-1}) AU $ and $ 254 (+150/-119)\times(V_T / 25,000 km s^{-1}) AU $, respectively. Similar values are obtained from the independent analysis of C (Table 1). From the derived size of the inner radius ($R_{in}$) of the accretion disk and using a relation for the last stable orbit ($ R_{in}\simeq3R_{sw} $) of a black hole with a Schwarzschild radius ($R_{sw}$), the mass of the SMBH in QSO 2237+0305 is estimated to be in the $1\sigma$ range $ 4.3 (+2.5/-2.3)\times(V_T /25,000 km s^{-1})10^{9} M_\odot $. This mass estimate is the largest one compared to known SMBHs (Gebhardt et al. \cite{gebh}). Our mass estimation for the SMBH of QSO2237 is almost compatible with Kochanek's recent result (\cite{koch}) and 10 or 100 times higher than Goicoechea's result (\cite{goi}), if we adopt $V_T=25,000kms^{-1}$. 

When adopting $ V_T=25,000kms^{-1} $, the estimated mass range for the microlenses is of the order of $10^{-2} M_\odot$, although one star with a mass $1 M_\odot$ is required to fit the HAE of C (see Table 1). Then, our mass estimation of microlenses in the Einstein Cross is almost consistent with the results by Kochanek (\cite{koch}). But, if we adopt a smaller transverse velocity, $V_T=6,000kms^{-1}$ (Kayser, Refsdal \& Stabell \cite{kayser}, Kayser \& Refsdal \cite {kayser89}), the estimated mass range is between $10^{-4} M_\odot \sim 10^{-3} M_\odot$. This is below the hydrogen-burning limit; it can thus represent a range of masses compatible with extra-galactic brown dwarves or free floating Jupiter-like planets. Refsdal \& Kayser (\cite {refsdal}) have previously anticipated that microlensing by small masses with a large source size ($R_E \leq R_S$) could explain the observed variability in the Einstein Cross. Our results may support their claim on the point that the source size is larger than the Einstein length of a mean mass of microlenses, since our source size estimation is slightly larger than the Einstein length of mean microlenses (Fig.1 \& Fig. 2). It also suggests the possibility that there can be numerous planetary masses in the distant lensing galaxy, if we adopt $ V_T=6,000kms^{-1}$. The exact masses of microlenses in the Einstein Cross still remain uncertain since there are very different estimations for the transverse velocity between $ 6,000kms^{-1} \sim 40,000kms^{-1} $. Accordingly, for example, the mass range corresponding to 100 days in the source plane is estimated between $ 0.0007 M_\odot \sim 0.033 M_\odot $ depending on the adopted transverse velocity. Though the exact masses of microlenses and the size of source are functions of the transverse velocity, the derived caustic size and the source size are invariant in time dimension or in normalized unit in LOHCAM (Fig. 1 \& Fig. 2).

Although we have firstly estimated the masses of microlenses in this galaxy under the assumption of the transverse velocity, it is not possible to derive or compare the expected surface mass density from our model. Because we must know both the mass and the spatial density of microlenses in order to derive the surface mass density, we can not estimate the surface mass density as we can not know the mean spatial density of microlenses from our modeling; we only estimate a few masses of microlenses, not their mean spatial density. Furthermore, the surface mass density in such a locally small region (see Fig. 1 \& Fig. 2) can be very fluctuated from the value expected in macro-lens models. However, it seems that the spatial density of microlenses in this system is not quite dense, otherwise our simple 5-microlens model could not reproduce the observed HAE variations. 

Our current estimate for the UV-continuum source size of the quasar measured in V-band (e.g diameter $ \simeq 3.94\times10^{16} (V_T/25,000kms^{-1}) cm $) is slightly larger than the estimated values by Shalyapin et al. (\cite{shal02}) and Yonehara (\cite{yonehara}), if we adopt $V_T=25,000kms^{-1}$. Although Wyithe et al. (\cite{wyithe2}) argued that the continuum source size of QSO 2237 would be much smaller than the Einstein length for the mean average mass of microlenses and neglected the low mass hypothesis by Refsdal \& Stabell (\cite{refsdal}), our results clearly show that it is possible to generate the strongest HAEs observed by OGLE with the combination of small size caustics compared to the source size. We also note that Wyithe et al.'s (\cite{wyithe}) prediction for the HAE in the Einstein Cross is completely failed; they predicted too much strong HAEs and it seems that they have based on too small source or too large caustics. If we assume a small transverse velocity, $V_T=6,000kms^{-1}$, they should be planetary masses. Then, our results are supporting Refsdal \& Kaiser (\cite{refsdal})'s idea that there can be a large amount of planetary masses, though the exact masses are still pending on the true transverse velocity in the source plane. Therefore, studying the transverse velocity in the source plane with a statistical analysis for the future OGLE light curves is very important so as to determine the exact size of the source and masses of microlenses in this system.

Considering two mutually perpendicular source track directions with respect to the galaxy center, we find that the whole HAEs of A are better modeled with the direction parallel to the shear direction (i.e. defined toward the galaxy center), whereas the whole HAEs of C are smoothly fitted with a track direction perpendicular to the local shear (Figs. 1 \& 2). Given the orthogonal positions of A \& C with respect to the galaxy center, these results may provide a rough, but consistent clue for the direction of the relative transverse motion of QSO 2237. Assuming that the random motions of the individual microlenses are negligible, we cautiously conclude that the direction of transverse motion of the source is more likely roughly parallel to the line connecting A \& B, instead of C \& D. This finding may imply that the true transverse velocity of the quasar can be larger than expected since the predicted observer's motion is parallel to the C-D direction (Witt \& Mao \cite{witt94}). 

Assuming that our interpretation of the source motion in the sky is correct, we can predict that such abrupt small fluctuations (cf. around day 1600 in Fig. 2) will be more frequent in image C, whereas image A will show smoother variations when microlensing HAE occurs. Also, the peak of HAE will be more sharp in C light curves as the predicted source's moving direction is perpendicular to the shear direction. Therefore, our current interpretation of the source motion should be further investigated on the basis of new data for the Einstein Cross. Especially, monitoring observation of the centroid shift in QSO2237 by the Space Interferometry Mission (SIM) is of great interest for us. If small mass lenses are dominant in this system, we could not expect to observe a significant centroid shift even during evident HAEs down to micro arcsecond scales. So the future SIM project may provide valuable information about the true transverse velocity and the masses of microlenses in this system. 

Our new technique, the LOHCAM, will play a key role for the future HAE analysis for extra-galactic microlensing events.

\begin{acknowledgements}We thank the OGLE team for making their data publicly available. This work was supported in part by PRODEX (Gravitational lens studies with HST),
by contract IUAP P5/36 ``P\^ole d'Attraction Interuniversitaire'' (OSTC, Belgium) and 
by the ``Fonds National de la Recherche Scientifique'' (Belgium). DWL would like to thank 
receiving a post-doc position and support from the 'Astrophysical Research Center for 
the Structure and Evolution of the Cosmos (ARCSEC)' in  Sejong university (Korea). DWL thanks C. Kochanek for very helpful comments on the new transverse velocity.
\end{acknowledgements}


\begin{thebibliography}{}


    \bibitem[1999]{agol} Agol E., Krolik J. 1999,
      ApJ, 524, 49


    \bibitem[2002]{alcalde} Alcalde D. et al. 2002,
      ApJ, 572, 729


    \bibitem[1979]{chang}  Chang K. \& Refsdal S. 1979,
      Nature, 282, 561
   
    \bibitem[1984]{chang84}  Chang K. \& Refsdal S. 1984,
      A\&A, 132, 168


    \bibitem[2000]{gebh}  Gebhardt K. et al. 2000,
      ApJ, 539, L13

    \bibitem[2003]{goi}  Goicoechea L J. et al.. 2003,
      A\&A, 397, 517 
 
    \bibitem[1988]{greiger}  Grieger B., Kayser R. \& Refsdal S.  1988,
      A\&A, 194, 54


    \bibitem[1985]{huchra} Huchra J. et al. 1985,
      AJ, 90, 691


    \bibitem[1986]{kayser}   Kayser R., Refsdal S. \& Stabell R. 1986,
      A\&A, 166, 36


    \bibitem[1989]{kayser89}   Kayser R., Refsdal S. 1989,
      Nature, 338, 745


    \bibitem[2004]{koch}   Kochanek C. 2004,
      ApJ, 605, 58


    \bibitem[1997]{lee} Lee D.-W. 1997,
      Master thesis, Kyunghee University, Korea. 

    \bibitem[1998]{lee2} Lee D.-W., Chang K. \& Kim S.J. 1998,
      Journal of Korean Astronomical Society (JKAS), 31, 27L

\bibitem[2003]{lee3} Lee D.-W. 2003,
      PhD thesis, University of Liege, Belgium. 

\bibitem[1999]{min} Mineshige, S., Yonehara, A. 1999,
       PASJ, 51, 497

    \bibitem[2004]{moreau} Moreau O., Libbrecht, C., Lee, D.W. \& Surdej, J. 2004,
     submitted to A\&A  

    \bibitem[1986]{pacz} Paczynski B. 1986,
      ApJ, 301, 503

    \bibitem[1992]{press} Press W. H., Teukolsky, S. A., Vetterling, W. T. \& Flannery, B. P. 1992,
     Numerical Recipes in C. Cambridge university press.


    \bibitem[1993]{refsdal}  Refsdal, S. \& Kayser, R.  1993,
      A\&A, 278, L5

    \bibitem[2001]{shal} Shalyapin, V. N. 2001,
      Astro-ph/0102384; Soviet Astron. Letter, 27, 150

    \bibitem[2002]{shal02} Shalyapin, V. N. et al. 2002,
      ApJ, 579, 127

    \bibitem[1998]{schmidt} Schmidt, Webster \& Lewis 1998,
      MNRAS, 295, 488

    \bibitem[1985]{sch} Schneider, P. et al 1985,
      AJ, 95, 1619

   \bibitem[1975]{thorne} Thorne, K. S. \& Price, R. H. 1975,
      ApJ, 195, L101

    \bibitem[1990]{wamb} Wambsganss, J., Paczynski, B. \& Schneider, P. 1990,
      ApJ, 358, L33

    \bibitem[1994]{wamb94}  Wambsganss, J. \& Paczynski, B.1994,
      AJ, 108, 1156

    \bibitem[1995]{witt} Witt, H., Mao, S. \& Schechter, P.L. 1995,
      ApJ, 443, 18

    \bibitem[1994]{witt94} Witt, H. \& Mao, S. 1994,
      ApJ, 429, 66

    \bibitem[2000]{wozniak} Wozniak, P. R. et al 2000,
      ApJ, 540, L65

    \bibitem[2000]{wyithe} Wyithe, J. S. B, Webster, R. L. \& Turner, E. L. 2000,
      MNRAS, 318, 1120

    \bibitem[2000]{wyithe2} Wyithe, J. S. B, Webster, R. L. \& Turner, E. L. 2000,
      MNRAS, 318, 762

    \bibitem[2001]{yonehara} Yonehara, A. 2001,
      ApJ,  548, L127

    \bibitem[1981]{young} Young, P. 1981,
      ApJ, 244, 756
  
\end{thebibliography}
\end{document}